\documentclass[prl,twocolumn,amsfonts,showpacs]{revtex4}
\usepackage{graphics,bm,natbib}
\begin{document}
\widetext
\title{Fingerprints of Spin-Orbital Physics in Crystalline O$_2$}
\author{I. V. Solovyev}
\email[Electronic address: ]{solovyev.igor@nims.go.jp}
\affiliation{Computational Materials Science Center, National Institute for Materials Science,
1-2-1 Sengen, Tsukuba, Ibaraki 305-0047, Japan
}
\date{\today}

\widetext
\begin{abstract}
The
alkali hyperoxide KO$_2$ is a molecular analog of strongly-correlated systems, comprising of
orbitally degenerate magnetic O$_2^-$ ions.
Using first-principles electronic structure calculations,
we set up an effective spin-orbital model for the low-energy
\textit{molecular} orbitals and argue that many anomalous properties of KO$_2$ replicate
the status of its orbital system
in various temperature regimes.
\end{abstract}



\maketitle

  Magnetic substances without $d$- or $f$-elements are
exotic. Several rare examples include $sp$-impurities in alkali-metal hosts~\cite{Papanikolaou},
CaB$_6$~\cite{Young}, nonstoicheometric CaO~\cite{Elfimov}, as well as some
carbon-based materials~\cite{Makarova}.
Besides them,
the solid oxygen
is an absolutely unique
magnetic system.
This is because O$_2$ is
the only elementary molecule, whose ground state
is different form the conventional spin singlet.
Therefore, if these molecules form a crystal (either by cooling or by pressurizing), it may become magnetic.
In fact, the antiferromagnetism of solid O$_2$ is well known~\cite{Hemert,Serra,Goncharenko},
and can be anticipated from the electronic structure of the single O$_2$ molecule,
having two unpaired electrons in a doubly-degenerate $\pi_g$ shell~\cite{Serra}.

\begin{figure}[h!]
\centering \noindent
\resizebox{8cm}{!}{\includegraphics{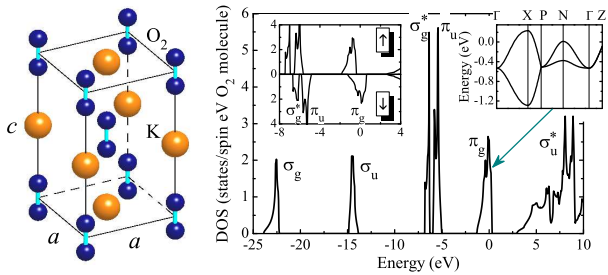}}
\caption{\label{fig.structure_DOS}
(Color online)
Tetragonal
phase of KO$_2$ (left)
and electronic structure in LDA (right).
Left inset shows results of spin-polarized calculations.
Right inset shows the
$\pi_{\rm g}$ bands near the
Fermi level (located at zero energy).}
\end{figure}

  Since
O$_2$ is a good oxidizer and can easily take an additional electron when
it brought in contact with alkali elements,
there is another way of making the crystalline O$_2$,
in a form of ionic crystals
(called ``alkali hyperoxides''), which are similar to NaCl,
but distorted due to elastic interactions involving
O$_2$ molecules.
One typical example is KO$_2$, which around the room temperature
forms the body-centered tetragonal (bct)
lattice (Fig.~\ref{fig.structure_DOS})~\cite{KanzigLabhart,Ziegler}.
The uniqueness of this situation is that O$_2$ has an extra electron
residing in the
doubly-degenerate $\pi_g$ shell, and KO$_2$ is the first molecular crystal,
where the magnetism not only exists, but interplays with the orbital degrees of freedom,
like in famous KCuF$_3$~\cite{KugelKhomskii} and LaMnO$_3$~\cite{KugelKhomskii,DE}.
Typically, such materials are known for their rich magnetic properties,
which depend
on the orbital state.

  This is clearly the case also for KO$_2$.
The Curie-Weiss temperature ($T_{CW}$), extracted from the behavior of
magnetic susceptibility ($\chi_m$) in two paramagnetic regions changes the sign at around
$200$ K~\cite{KanzigLabhart},
indicating at the change of character of intermolecular interactions,
from antiferromagnetic (AFM) to ferromagnetic (FM) with decrease of the temperature ($T$).
The role of the crystal distortion in this region is somewhat controversial.
On the one hand, there are some indications that the actual symmetry is lower than
bct~\cite{KanzigLabhart,Ziegler}.
On the other hand, no profound change of the magnetic behavior has been observed.
For example, a sizable orbital contribution to the magnetization persists down to $13$~K.
Therefore, even if exists, the crystal distortion does not seem to
fully control the magnetic behavior of KO$_2$ in the paramagnetic region.
If so, what causes the change of
$T_{CW}$?
Another interesting point is that
below $13$~K the situation is completely different and
the behavior of KO$_2$ is strongly affected by frozen-in rotations of O$_2$ molecules, which quenche the
orbital magnetization and stabilize an AFM order.
Nevertheless, the transition temperature is surprisingly low.
Moreover,
the AFM order can be broken up by a magnetic field. The transition is accompanied by structural changes,
and
the effect is called ``magnetogyration''~\cite{magnetogyration}.

  The purpose of this work is to establish a theoretical basis for understanding
the fascinating magnetic properties of
KO$_2$ and other
alkali hyperoxides.
Using
first-principles electronic structure calculations,
we set up
an effective spin-orbital model for the low-energy molecular states.
Then, we show that the magnetic phase diagram of KO$_2$
replicates the status of its orbital system, which evolves
(under heating) from
a quenched state, to a region of
relativistic spin-orbital correlations, and finally to the region of
independent spin and orbital disorder.

  The O$_2$ molecule appears to be the building block not only of the crystal,
but also of the electronic structure of KO$_2$. The hybridization within a single
molecule is so strong that it leads to the splitting and formation of quite
distinguishable molecular levels. The interaction between
molecules is considerably weaker, so that the molecular orbitals form a group
of narrow nonoverlapping bands
(Fig.~\ref{fig.structure_DOS}, we use the experimental lattice
parameters reported in Ref.~\cite{Ziegler}). Thus, there is a
clear analogy with atomic limit in the theory of strongly-correlated systems~\cite{KugelKhomskii},
except that now
the \textit{molecular orbitals} play the
same
role as atomic orbitals in a conventional case.

  According to electronic structure calculations in the
local-(spin)-density approximation [L(S)DA], the doubly-degenerate
$\pi_g$ band located near the Fermi level is formed by antibonding
molecular $p_x$ and $p_y$ orbitals (Fig.~\ref{fig.structure_DOS})~\cite{Priya}.
The $\pi_g$ bandwidth is comparable with
the exchange spin-splitting ($\Delta_{\rm ex}$$\sim$$1$ eV).
Hence, the system is half-metallic, and its magnetic properties can be understood
in terms of FM double exchange ($D$) and
AFM superexchange ($S$) interactions,
which can be expressed through the first and second moments of
occupied density of states for the $\downarrow$-spin $\pi_g$ band as
$J^D$$=$$-$$m^{(1)}/2z$ and $J^S$$=$$-$$m^{(2)}/z\Delta_{\rm ex}$, where
$z$ is the coordination number~\cite{DE}.
Then, the Curie temperature will also consist of two parts,
$T_C$$=$$T_C^D$$+$$T_C^S$, which
can be evaluated in the mean-field approximation~\cite{deGennes}.
This yields
$T_C^D$$=$$-$$4m^{(1)}/15 k_B \approx 882$ K
and $T_C^S$$=$$-$$m^{(2)}/3 k_B \Delta_{\rm ex} \approx - 558$ K.
Thus, $T_C \approx 324$ K, and KO$_2$ is expected to be a good ferromagnet,
being in straight
contrast with the experimental data. The root of the problem is the
Coulomb correlations in the narrow $\pi_g$ band, which are
greatly oversimplified in LSDA.

  In order to treat these effects rigorously, we derive an effective
Hubbard-type model for the $\pi_g$ band, starting from the LDA band structure:
\begin{equation}
{\cal H} = \sum_{ij} \sum_{\alpha \beta}
h_{ij}^{\alpha \beta} c^\dagger_{i \alpha} c^{\phantom{\dagger}}_{j \beta}
+ \sum_i \sum_{\alpha \beta \gamma \delta}
U_{\alpha \beta \gamma \delta}
c^\dagger_{i \alpha} c^\dagger_{i \beta} c^{\phantom{\dagger}}_{i \gamma} c^{\phantom{\dagger}}_{i \delta}.
\label{eqn:Hubbard}
\end{equation}
The method
has been explained in details in Ref.~\cite{PRB06}.
The basic difference here is that each lattice-point ($i$ or $j$)
corresponds to an O$_2$ molecule (rather than to a single atomic
site). Each Greek symbol stands for a combination ($m,s$) of
orbital ($m$$=$ $p_x$ or $p_y$)
and
spin
($s$$=$ $\uparrow$ or $\downarrow$)
indices. Each orbital index refers to the \textit{antibonding molecular p-orbital}
of either $x$ ($m$$=$$1$) or $y$ ($m$$=$$2$) symmetry.
The off-diagonal (with respect to $i$ and $j$) elements of $h_{ij}^{\alpha \beta}$
stand for transfer integrals, which do not depend on spin indices: i.e.,
$\hat{h}_{ij}$$\equiv$$\| h_{ij}^{\alpha \beta} \|$$=$$\| t_{ij}^{mm'} \| \delta_{ss'}$
for any combination of $\alpha$$=$$(m,s)$ and $\beta$$=$$(m',s')$.
They are derived
from the LDA band structure
using an \textit{exact} version of the downfolding method~\cite{PRB06,preprint06}.
The results are summarized in Table~\ref{tab:Table}, up to fourth
nearest neighbor.
Other parameters are considerably smaller.
The diagonal matrix elements,
$h_{ii}^{\alpha \beta}$$=$$\langle \alpha | \xi \hat{\bf L} \cdot \hat{\bf S} | \beta \rangle$,
stand for the spin-orbit interaction (SOI) is a single molecule
($\xi \approx 34$ meV) and the crystal field, if the lattice distortion is applied.

\begin{table}[h!]
\caption{\label{tab:Table} Transfer integrals and parameters of
Heisenberg model (in meV) for undistorted bct phase.
The nonequivalent vectors separating two O$_2$ molecules are denoted as
${\bf b}_1$$=$$(0,a,0)$,
${\bf b}_2$$=$$(\frac{a}{2},\frac{a}{2},\frac{c}{2})$,
${\bf b}_3$$=$$(a,a,0)$, and
${\bf b}_4$$=$$(0,0,c)$. Other matrix elements $t_{ij}^{mm'}$ are obtained using
symmetry operations and permutability of orbital indices, $t_{ij}^{12}$$=$$t_{ij}^{21}$.}
\begin{ruledtabular}
\begin{tabular}{crrrrrr}
vector              & $t^{11}_{ij}$  & $t^{12}_{ij}$ & $t^{22}_{ij}$ & $J^\parallel_{ij}$ & $J^\perp_{ij}$  &  $\bar{J}_{ij}$ \\
\colrule
${\bf b}_1$         & $-$$1$    &    $0$   & $106$    & $-$$0.44$     &  $-$$1.09$ &  $-$$0.47$ \\
${\bf b}_2$         & $-$$51$   &  $-$$98$ & $-$$51$  &  $1.13$       &  $-$$1.03$ &  $-$$1.03$ \\
${\bf b}_3$         & $-$$11$   &  $-$$11$ & $-$$11$  & $-$$0.02$     &  $-$$0.04$ &  $-$$0.02$ \\
${\bf b}_4$         & $-$$13$   &    $0$   & $-$$13$  & $-$$0.06$     &  $-$$0.06$ &  $-$$0.01$ \\
\end{tabular}
\end{ruledtabular}
\end{table}

  The matrix of screened Coulomb interactions in the $\pi_g$ band has been
computed in the two steps.
First, we derive the
interaction parameters between \textit{atomic} $p$-orbitals,
using the constraint-LDA method.
It enable us to obtain the following parameters of
intraatomic Coulomb interaction $u \approx 11.37$ eV,
interatomic intramolecular Coulomb interaction $v \approx 2.52$ eV,
and intraatomic exchange interaction $j \approx 2.30$ eV.
Then, we take into account the screening of Coulomb interactions between
$p$-orbitals in the
$\pi_g$ band by other bands, which are constructed from the same
$p$-orbitals (i.e., the $\pi_u$ band, which is the bonding
combination of $p_x$- and $p_y$-orbitals, as well as the
$\sigma_g^*$ and $\sigma_u^*$ bands, which have a strong
weight of $p_z$-orbitals).
This part is done in the random-phase approximation,
which starts with interaction parameters obtained in constraint-LDA.
Then, we are able to derive the interaction parameters
between \textit{molecular}
$p$-orbitals in the $\pi_g$ band:
the intraorbital Coulomb interaction $U \approx 3.66$ eV
and the exchange interaction $J \approx 0.62$ eV.
The interorbital Coulomb interaction $U'$
is related with $U$ and $J$ by the identity $U'$$=$$U$$-$$2J$.
These parameters define the whole matrix
$\hat{U}$$=$$\| U_{\alpha \beta \gamma \delta} \|$
of screened Coulomb interactions in the $\pi_g$ band.

  Since
any of $t_{ij}^{mm'}$ is smaller than $U$ by at least one order of magnitude,
all transfer integrals can be treated
as a perturbation,
starting with
isolated molecular orbitals.
Then,
it is convenient to use the \textit{hole representation}.
The advantage is that there is only one hole state associated with
each molecular site. Therefore, if molecules were fully isolated,
the holes would not interact with each other, and the ground state
would be a single Slater determinant.
By denoting the hole-orbitals associated with $i$ and $j$
as $\alpha_i$ and $\alpha_j$, and constructing the two-hole
determinant
$G(\alpha_i, \alpha_j)$$=$$\frac{1}{\sqrt{2}}\{\alpha_i(1) \alpha_j(2)$$-$$\alpha_j(1) \alpha_i(2) \}$,
one can find the energy gain caused by virtual hoppings
in each bond~\cite{condmat06,SEmultiplet}:
\begin{widetext}
\begin{equation}
{\cal T}(\alpha_i,\alpha_j) = - \left\langle G(\alpha_i,\alpha_j)
\left| \hat{h}_{ij} \left( \sum_M \frac{{\hat{\cal P}}_j|j M \rangle \langle j M|
{\hat{\cal P}}_j}{E_{jM}} \right) \hat{h}_{ji} +
(i \leftrightarrow j) \right| G(\alpha_i,\alpha_j) \right\rangle,
\label{eqn:egain}
\end{equation}
\end{widetext}
where $E_{jM}$ and $|jM \rangle$ stand for eigenvalues and eigenstates of
excited two-hole configurations at the site $j$, and $\hat{\cal P}_j$ is a projector
operator, which enforces the Pauli principle and prevent any hoppings onto
$\alpha_j$~\cite{condmat06}. Eq.~(\ref{eqn:egain}) incorporates some
intramolecular correlations beyond the Hartree-Fock approximation, which are included
into the multiplet structure of the excited states.

  First, we discuss the properties of
the bct phase of
KO$_2$ in an intermediate-temperature region,
where the crystal distortion is small and
$k_BT$$\ll$$\xi$ (note that $\xi/k_B$$\approx 395$ K),
so that the splitting of
the molecular $\pi_g$ levels is largely controlled by the SOI,
which provides the natural basis for the hole orbitals $\{ \alpha_i \}$.
In this case,
each $| \alpha_i \rangle$
can be presented as a linear combination of
only
two orbitals, $|p_+$$\uparrow \rangle$ and $|p_-$$\downarrow \rangle$,
which are split off by the SOI, and where
$| p_\pm \rangle$$=$$\mp$$\frac{1}{\sqrt{2}}( | p_x \rangle$$ \pm$$  i | p_y \rangle$$ )$
are the complex harmonics.
Now, the spin and orbital variables are rigidly bound, and
the problem can be reformulated in terms of the \textit{pseudospin} states:
$| \tau_x^\pm \rangle$$=$$\frac{1}{\sqrt{2}}( |p_+$$\uparrow \rangle$$\pm$$|p_-$$\downarrow \rangle )$,
$| \tau_y^\pm \rangle$$=$$\frac{1}{\sqrt{2}}( |p_+$$\uparrow \rangle $$\pm$$ i|p_-$$\downarrow \rangle )$,
$| \tau_z^+ \rangle$$=$$|p_+$$\uparrow \rangle$, and $| \tau_z^- \rangle$$=$$|p_-$$\downarrow \rangle$,
which allow us to make a formal mapping of
energies (\ref{eqn:egain}) onto the \textit{anisotropic} Heisenberg model for the bct lattice
and the pseudospin $1/2$:
$$
\hat{H} = - \frac{1}{2} \sum_{ij}
\left\{
\left( \hat{\tau}_{ix} \hat{\tau}_{jx} + \hat{\tau}_{iy} \hat{\tau}_{jy} \right) J^\perp_{ij} +
\hat{\tau}_{iz} \hat{\tau}_{jz} J^\parallel_{ij}
\right\},
$$
where
$\hat{\tau}_x$, $\hat{\tau}_y$, and $\hat{\tau}_z$ are the $2$$\times$$2$ Pauli matrices
in the basis of $|p_+$$\uparrow \rangle$ and $|p_-$$\downarrow \rangle$ orbitals.
The parameters,
calculated as
$2J^\perp_{ij}$$=
{\cal T}(\tau_{ix}^+,\tau_{jx}^-)$$-$${\cal T}(\tau_{ix}^+,\tau_{jx}^+)$$=$$
{\cal T}(\tau_{iy}^+,\tau_{jy}^-)$$-$${\cal T}(\tau_{iy}^+,\tau_{jy}^+)$ and
$2J^\parallel_{ij}$$=
{\cal T}(\tau_{iz}^+,\tau_{jz}^-)$$-$${\cal T}(\tau_{iz}^+,\tau_{jz}^+)$,
are summarized
in Table~\ref{tab:Table}.
The direct exchange interactions between O$_2$ molecules are
considerably weaker and can be neglected~\cite{Hemert}.
The interlayer coupling $J^\parallel$
operating
between neighboring molecules
separated by ${\bf b}_2$
stabilizes the FM ground state.
Hence, the pseudospin moments are parallel to the $z$ axis.
Other interactions are antiferromagnetic
and frustrated on the bct lattice.
The expectation values of the magnetic moments are given by the following
identities:
$\langle \tau_x^\pm | \hat{L}_x$$+$$2\hat{S}_x | \tau_x^\pm \rangle$$=$$
\langle \tau_y^\pm | \hat{L}_y$$+$$2\hat{S}_y | \tau_y^\pm \rangle$$=$$0$
and
$\langle \tau_z^\pm | \hat{L}_z$$+$$2\hat{S}_z | \tau_z^\pm \rangle$$=$$\pm$$2$:
i.e., the finite magnetization is allowed only
along the $z$-axis, while it is totally quenched in the perpendicular ($x$ and $y$)
directions. The same matrix elements define the $g$-tensor (e.g., in $\chi_m$).

\begin{figure}[h!]
\centering \noindent
\resizebox{8cm}{!}{\includegraphics{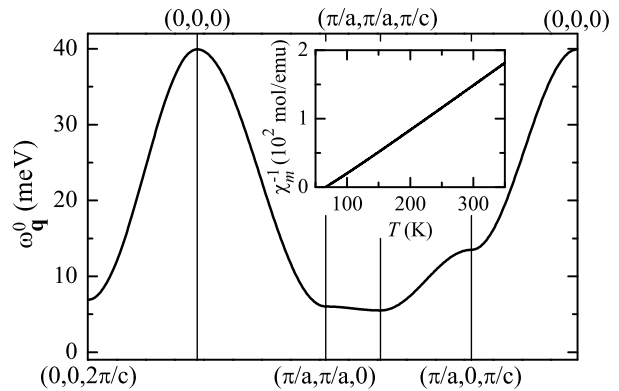}}
\caption{\label{fig.magnons}
The unrenormalized spin-wave dispersion of anisotropic Heisenberg model:
$\omega^0_{\bf q}$$=$$2( J^\parallel_0$$-$$J^\perp_{\bf q} )$.
The inset shows the inverse magnetic susceptibility
in the paramagnetic region, as obtained in the renormalized
spin-wave theory.}
\end{figure}

  Then, the magnetic properties can be easily calculated
using renormalized spin-wave (SW) theory~\cite{Tyablikov}. The idea is to extend the
conventional formalism to finite temperatures by replacing
in the expression for the SW frequency
the value of
local
magnetic moment (in our case, the pseudospin moment) by its thermal average
$\langle \hat{\tau}_z \rangle$:
$\omega_{\bf q}$$=$$2 \langle \hat{\tau}_z \rangle
( J^\parallel_0$$-$$J^\perp_{\bf q} )$$+$$2b$,
where $J^\parallel_0$$=$$\sum_j J^\parallel_{ij}$ and
$J^\perp_{\bf q}$ is the Fourier transform of $\{ J^\perp_{ij} \}$.
On the other hand, $\langle \hat{\tau}_z \rangle$ is expressed through
the averaged number of excited spin-waves,
yielding the equation
$\langle \hat{\tau}_z \rangle
\sum_{\bf k} \coth \omega_{\bf k}/2 k_B T = 1$,
which is solved self-consistently
for each $T$ and
the external field $b$, acting on $ \hat{\tau}_z $.
The expression for the Curie temperature,
$(k_B T_C)^{-1}$$=$$ \sum_{\bf q} (J^\parallel_0$$-$$J^\perp_{\bf q})^{-1}$,
can be derived by assuming $b$$=$$\langle \hat{\tau}_z \rangle$$=$$0$
or from the divergence of
$\chi_m$$=$$4 \mu_B^2 \langle \hat{\tau}_z \rangle/b$.
The spectrum of SW excitations itself is rather remarkable (Fig.~\ref{fig.magnons}).
All excitations are gapped.
The gap
in the Brillouin zone center is proportional to $\xi$,
which is large on the temperature scale. However, all low-energy excitations
correspond to the zone boundary, where $\omega_{\bf q}$ is largely reduced
by AFM
interactions $J^\perp_{ij}$.
Thus, the
FM order is expected only at
$T_C$$\approx 66$ K, which is largely reduced in comparison with $\xi/k_B$
(and, apparently, can be further reduced by small lattice distortions,
which are seen in KO$_2$~\cite{KanzigLabhart,Ziegler}).
The renormalizes SW theory also predicts some small deviation from the
Curie-Weiss law in the paramagnetic region, and the effective moment ($\mu_{\rm eff}$),
derived from the slope of $\chi_m^{-1}$, will generally depend on $T$.
Indeed, $\mu_{\rm eff}$ varies
from $2$ $\mu_B$ at large $T$ till
$2.17$ $\mu_B$ near $T_C$.
This may help to explain some deviations of $\mu_{\rm eff}$ from its nominal value~\cite{KanzigLabhart}.

  In the high-temperature region, $k_BT$$\gtrsim$$\xi$, the thermal fluctuations will
eventually destroy the coupling between spin and orbital degrees of freedom, which now can
be treated as independent variables. Then, the
\textit{spin} Hamiltonian can be derived by averaging the pair interactions
(\ref{eqn:egain}) over
\textit{orbital} variables~\cite{KugelKhomskii}.
Practically, for each projection of spin, one can consider the
whole family of possible hole-orbitals,
$| \alpha_i^{\uparrow,\downarrow} \rangle$$=
$$\cos \theta_i | p_x$$\uparrow,\downarrow \rangle$$+
$$\sin \theta_i e^{i \phi_i} | p_y$$\uparrow,\downarrow \rangle$
(where $0$$\le$$\theta_i$$\le$$\pi$ and $0$$\le$$\phi_i$$\le$$2\pi$),
and to average (numerically) each ${\cal T}(\alpha_i^{\uparrow,\downarrow},\alpha_j^{\uparrow,\downarrow})$
over all combinations of
$(\theta_i,\phi_i)$ and $(\theta_j,\phi_j)$.
The simplest approximation,
which is justified for large $T$, is to
assume that all combinations are equivalent
and to neglect the spacial correlations between orbitals.
Obviously, the spin system will be isotropic, and the
parameters of Heisenberg model can be found as
$2 \bar{J}_{ij}$$=$$\bar{\cal T}_{ij}^{\uparrow \downarrow}$$-$$\bar{\cal T}_{ij}^{\uparrow \uparrow}$,
where $\bar{\cal T}_{ij}^{\uparrow \downarrow}$ and $\bar{\cal T}_{ij}^{\uparrow \uparrow}$
are the averaged values of ${\cal T}(\alpha_i^\uparrow,\alpha_j^\downarrow)$
and ${\cal T}(\alpha_i^\uparrow,\alpha_j^\uparrow)$, respectively.
All parameters are antiferromagnetic (Table~\ref{tab:Table}),
which naturally explain the sign of $T_{CW}$
in the high-temperature region~\cite{KanzigLabhart}.
The absolute value of $T_{CW}$ can be estimated from the behavior of
spin susceptibility ($\chi_s$),
in the framework of renormalized SW theory.
Note that $\chi_s$ alone
should provide a good estimate for the total $\chi_m$ in the perpendicular
$xy$-plane, where the orbital magnetization is  quenched.
This yields $T_{CW}$$\approx -$$100$ K,
which was derived from the linear interpolation of $\chi_s^{-1}$
at large $T$.

  Thus,
the existence of several paramagnetic segments with different $T_{CW}$'s
can be related with
a gradual increase of on-site correlations
between spin and orbital degrees of freedom,
driven by the SOI \textit{in the bct phase}.
In the high-temperature region, the spin and orbital variable are totally decoupled,
that explains the AFM character of intermolecular interactions.
On the other hand, the strong spin-orbital correlations in the intermediate-temperature regime
would lead to a ferromagnetism.

  The crystal distortion plays a sizable role only at very low $T$, as it is
clearly manifested in the quenching of orbital moments and
the
observed AFM
order~\cite{KanzigLabhart}.
Although exact details of the low-temperature structure are unknown,
a number of experimental data speak
in favor of uniform rotations of O$_2$ molecules around
one of tetragonal
axes as the main distortion.
Certainly, such a distortion will lift the degeneracy of molecular $p_x$ and
$p_y$ levels, and lead to a FM orbital order,
which typically coexists with
AFM
correlations between the spins~\cite{KugelKhomskii}.
However, the situation is not so simple.
Let us consider a distorted structure, which was obtained
by the rigid clockwise rotation of O$_2$ molecules
around the $y$-axis by $30^\circ$ (the experimental value~\cite{KanzigLabhart,Ziegler})
in the fixed tetragonal frame, and derive parameters of the model
Hamiltonian (\ref{eqn:Hubbard}) for the (deformed) $\pi_g$ band.
The obtained ``crystal-field splitting'' of molecular $\pi_g$ levels is about $290$ meV,
which
exceeds $\xi$ by nearly one order of magnitude
and
stabilizes the $p_y$-orbital (opposite to the assumption made in the
phenomenological theory of magnetogyration~\cite{magnetogyration}).
Generally, the transfer integrals are specified by all three
parameters $(t^{11},t^{12},t^{22})$~\cite{comment1}.
For the neighboring O$_2$ molecules, they are
$(-$$20,0,74)$, $(-$$25,0,2)$, $(37,-$$29,-$$67)$, and $(-$$43,35,-$$21)$ meV, for
${\bf b}^{\phantom{*}}_1$$=$$(0,a,0)$, ${\bf b}^*_1$$=$$(a,0,0)$,
${\bf b}^{\phantom{*}}_2$$=$$(\frac{a}{2},\frac{a}{2},\frac{c}{2})$, and
${\bf b}^*_2$$=$$(-$$\frac{a}{2},\frac{a}{2},\frac{c}{2})$, respectively.
Thus,
the transfer integrals are strongly anisotropic, both in and between the tetragonal
$xy$-planes.
Since the SOI is no longer important,
the energies (\ref{eqn:egain}) can be mapped onto
rotationally
invariant ($J^\perp_{ij}$$=$$J^\parallel_{ij}$) Heisenberg model.
The natural choice of the hole orbitals,
$| p_y$$\uparrow \rangle$ or $| p_y$$\downarrow \rangle$, is dictated by
the crystal-field splitting.
This yields the following interaction parameters
$J_{ij}$$=$ $-$$0.77$, $0$, $-$$0.59$, $4$, $-$$0.09$, and $-$$0.01$ meV
between O$_2$ molecules
separated by the vectors
${\bf b}^{\phantom{*}}_1$, ${\bf b}^*_1$,
${\bf b}^{\phantom{*}}_2$, ${\bf b}^*_2$, ${\bf b}^{\phantom{*}}_3$,
and ${\bf b}^{\phantom{*}}_4$, respectively.
They form quasi-two-dimensional networks, where interactions in the planes
perpendicular to $[{\bf b}^{\phantom{*}}_1$$\times$${\bf b}^{\phantom{*}}_2]$
are much stronger than those between the planes.
This suppresses any long-range magnetic order~\cite{MerminWagner}. Moreover,
the AFM interactions in each plane are frustrated.
Therefore, the N\'{e}el temperature ($T_{\bf Q}$) is expected to be small.
Indeed, from the mean-field analysis
of the Heisenberg model we conclude that the ground state is a (nearly AFM) spin-spiral
with ${\bf Q}$$\approx (0,\frac{3.77}{a},\frac{0.02}{c})$.
Then, the renormalized SW theory yields $T_{\bf Q}$$\approx 11$ K, being in good agreement
with the experiment~\cite{KanzigLabhart}.
The first-order transition to the FM bct phase in a magnetic field
is also anticipated.

  In summary,
KO$_2$ is the new molecular analog of correlated spin-orbital systems.
The geometry of \textit{molecular} orbitals adds new functionalities
into this classical problem.
There is also a number of open questions.
Particularly,
any
quantitative theory of magnetogyration is missing.
Why do the O$_2$ molecules tend to rotate at low $T$?
Why do their orientation is different in KO$_2$ and NaO$_2$?
These problems prompt further research.

  I wish to thank P.~Mahadevan for drawing my attention to KO$_2$,
as well as M.~Boero and Y.~Motome for useful discussions.
This work is partly supported by Grant-in-Aids
for Scientific Research in Priority Area ``Anomalous Quantum Materials''
from the Ministry of Education, Culture, Sports, Science and Technology of Japan.


\begin{thebibliography}{10}
\bibitem{comment1}
Although the centers of O$_2$ molecules are located in the high-symmetry
positions, the atoms themselves are shifted off
the tetragonal axes, and
the conventional Slater-Koster parametrization of $t_{ij}^{mm'}$ is not applicable.
\end{thebibliography}
\end{document}